\documentclass[reprint,superscriptaddress,aps,prb,twocolumn]{revtex4-1}
\usepackage[toc,page]{appendix}
\usepackage{braket}
\usepackage{amsmath}
\usepackage{amsfonts}
\usepackage{amssymb}
\usepackage{graphicx}
\usepackage[colorlinks,linkcolor=blue,anchorcolor=blue,citecolor=blue,urlcolor=black]{hyperref}
\usepackage{mathrsfs}
\usepackage{dcolumn}
\usepackage{bm}
\usepackage{epsfig}
\usepackage{placeins}
\usepackage{textcomp}
\usepackage{bbold}
\usepackage{float}
\usepackage{verbatim}
\renewcommand{\figurename}{\textbf{Figure}}
\begin{document}

\title{Enhanced sensitivity operation of an optical gyroscope near an exceptional point} 
\author{Yu-Hung Lai$^{1\ast}$, Yu-Kun Lu$^{1,2\ast}$, Myoung-Gyun Suh$^{1\ast}$, Kerry Vahala$^{1\dagger}$ \\
$^1$T. J. Watson Laboratory of Applied Physics, California Institute of Technology, Pasadena, California 91125, USA \\
$^{2}$State Key Laboratory for Mesoscopic Physics and Collaborative Innovation Center of Quantum Matter, School of Physics, Peking University, Beijing 100871, People’s Republic of China \\
$^\ast$These authors contributed equally to this work. \\ 
$^\dagger$vahala@caltech.edu}
% \date{\today}
\maketitle

\noindent\textbf{Exceptional points (EPs) are special spectral degeneracies of non-Hermitian Hamiltonians governing the dynamics of open systems. At the EP two or more eigenvalues and the corresponding eigenstates coalesce \cite{el2018non,feng2017non,miri2019exceptional}. Recently, it has been proposed that EPs can enhance the sensitivity of optical gyroscopes \cite{ren2017ultrasensitive,PhysRevA.96.033842}. Here we report measurement of rotation sensitivity boost by over 4$\times$ resulting from operation of a chip-based stimulated Brillouin gyroscope near an exceptional point. A second-order EP is identified in the gyroscope and originates from the dissipative coupling between the clockwise and counterclockwise lasing modes. The modes experience opposing Sagnac shifts under application of a rotation, but near the exceptional point new modal admixtures dramatically increase the Sagnac shift. Modeling confirms the measured enhancement. Besides the ability to operate an optical gyroscope with enhanced sensitivity, this result provides a new platform for study of non-Hermitian physics and nonlinear optics with precise control.}

High-Q optical microresonators have received considerable attention as sensors across a wide range of applications including biomolecule \cite{vollmer2008whispering,Lu2011,vollmer2012review} and nanoparticle detection \cite{zhu2010chip}, temperature measurement \cite{xu2016phone}, and rotation measurement \cite{li2017microresonator,liang2017resonant,maayani2018flying,khial2018nanophotonic,gundavarapusub}. In recent years, a new approach to enhance the sensitivity of microresonator sensors using the physics of exceptional points is being studied \cite{ren2017ultrasensitive,PhysRevA.93.033809,PhysRevA.96.033842,PhysRevLett.112.203901,PhysRevA.93.033809,PhysRevLett.117.110802,hodaei2017enhanced,chen2017exceptional}. Traditionally, for precise sensing, a perturbation to an optical microcavity (or to its reference frame as in the case of a gyroscope) introduces either a linewidth change, a frequency shift, or a frequency splitting of a resonance that monotonically changes with the strength of the perturbation. However, operation of these systems near an exceptional point changes this situation by introduction of a square-root dependence into the transduction that can boost the sensor's ability to transduce perturbations \cite{PhysRevA.93.033809}.

\begin{figure*}
    \centering
    \includegraphics[width=2.05\columnwidth]{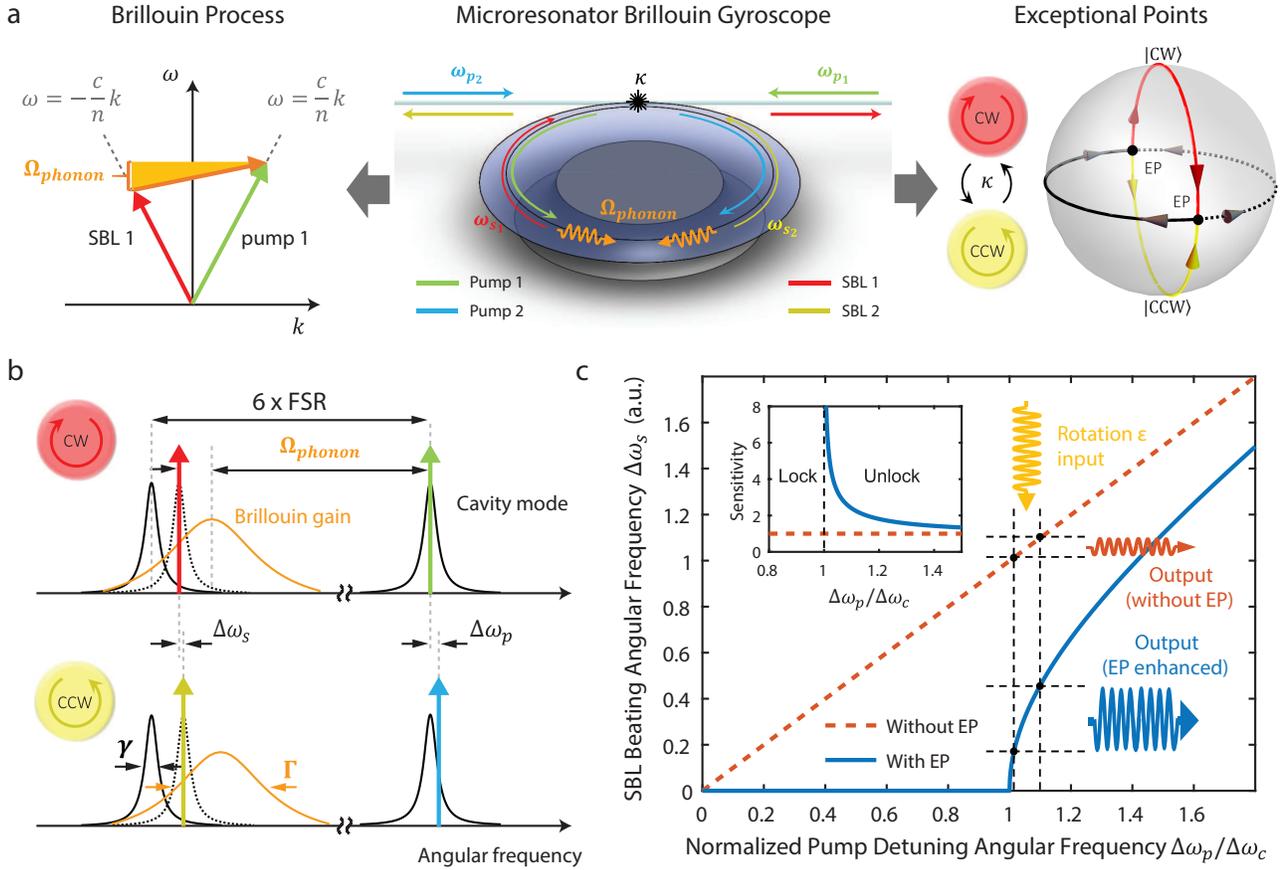}
    \caption{\textbf{EP-enhanced microresonator-based Brillouin gyroscope. }\textbf{a}, The dual stimulated Brillouin laser process in a microresonator. Center: The green (blue) solid curve represents pump 1 (pump 2) with angular frequency $\omega_{p_1}$ ($\omega_{p_2}$) and the red (yellow) solid curve represents SBL 1 (SBL 2) with angular frequency $\omega_{s_1}$ ($\omega_{s_2}$). The orange wavy line represents the acoustic phonons with angular frequency $\Omega_{\textrm{phonon}}$. Left: The Brillouin energy and the momentum conservation constraints (phase matching) are illustrated for scattering of a pump wave into a Stokes wave. Right: Because of resonator imperfections including the fiber taper, the CW and CCW modes experience dissipative coupling at rate $\kappa$. This coupling creates eigenmodes that map to a Bloch sphere containing dual EPs (black dots). The trajectories on the Bloch sphere show the evolution of two eigenmodes (red for SBL1 and yellow for SBL2) when the pump detuning decreases from $+\infty$ to $-\infty$. The low-loss and high-loss eigenmodes inside the locking zone are plotted in solid and dashed black curves, respectively (see Methods for additional discussion). \textbf{b}, Efficient laser action requires that each Stokes mode (black with linewidth $\gamma$ and separated from the pump by a multiple of the cavity FSR) lies within the Brillouin gain band (orange with linewidth $\Gamma$) which, through the phase matching condition, is shifted relative to the pump by $\Omega_{\textrm{phonon}}=4\pi nc_{s}/\lambda_p$ (refractive index $n$, speed of sound in silica $c_{s}$ and pump wavelength $\lambda_p$). In this work, the FSR is $\sim$1.8 GHz so that 6$\times$FSR approximately matched the Brillouin shift. Dispersion from the Brillouin gain pulls the Stokes lasing modes by different amounts towards the gain center on account of the difference $\Delta \omega_p$ in pump angular frequencies. \textbf{c}, The blue solid curve (red dashed curve) shows the dependence of the dual-SBL beating angular frequency $\Delta\omega_s$ versus the normalized pump detuning frequency $\Delta\omega_p/\Delta\omega_c$ for $\kappa \not=0$ ($\kappa=0$) as per Eq. \ref{Eq4}. The yellow wavy arrow represents the input rotation signal, while the red dotted and blue solid wavy arrows represent the output signal with and without EP, respectively. The inset shows the $\kappa \not=0$ sensitivity normalized to the $\kappa =0$ sensitivity, indicating the enhancement near the EP.}
\label{Fig1}
\end{figure*}

\begin{figure*}
\centering
\includegraphics[width=2.05\columnwidth]{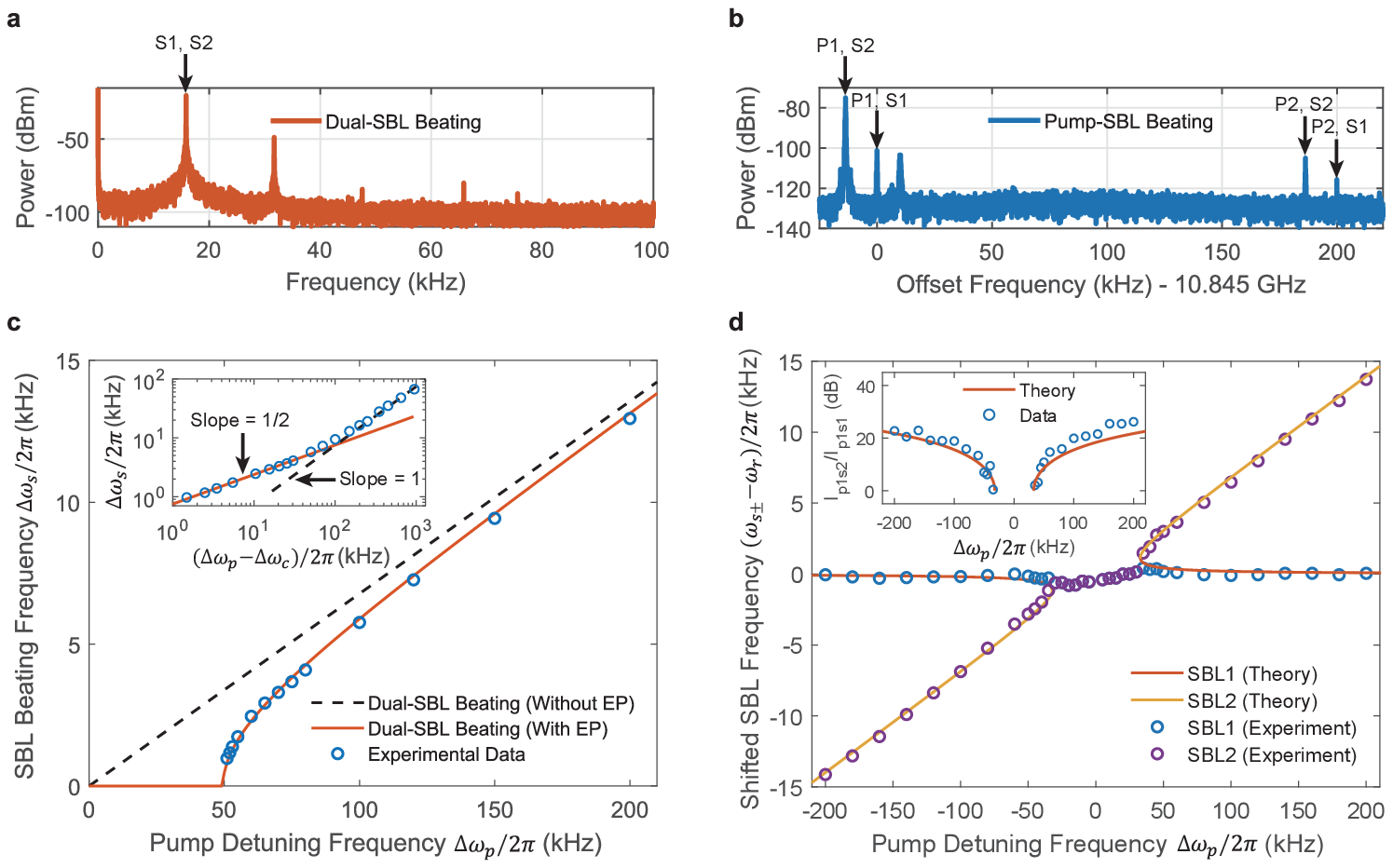}               
\caption{\textbf{Measurement of the eigenmode properties.}
\textbf{a}, Typical measured dual-SBL beating spectrum. \textbf{b}, Typical pump-SBL beating spectrum with frequency axis shifted approximately $10.845$ GHz to center the pump1-SBL1 beating peak. The individual pump-SBL beating peaks are identified. \textbf{c}, Measured dual-SBL beating frequency versus pump detuning frequency (blue circles). Red solid curve is fitting ($\gamma/\Gamma=0.073$ and $\kappa=1.80$kHz) and black dotted line corresponds to $\kappa = 0$ ($\gamma/\Gamma=0.073$). The data have a slope 1/2 (slope 1) near (away from) the EP in the log-log plot provided in the inset. This data used another mode with larger $\kappa$ compared to panel d. \textbf{d}, Measured shifted frequencies of the two SBLs $(\omega_{s\pm}-\omega_r)/2\pi$ versus pump detuning frequency. Theoretical values of $(\omega_{s+}-\omega_r)/2\pi$ and $(\omega_{s-}-\omega_r)/2\pi$ with $\gamma/\Gamma=0.076$ and $\kappa=1.23$kHz are plotted as red and dashed-yellow lines, respectively. The experimental data of the shifted SBL1 (SBL2) frequency is shown as blue (purple) circles. The inset shows the measured power ratio of CCW components of the lasing modes (blue circles) obtained by analysis of spectral components in (b) and agrees reasonably well with the theoretical prediction (red solid curve).}
\label{Fig2}
\end{figure*}

In this work, we experimentally and theoretically demonstrate the existence of EPs in a microresonator-based laser gyroscope, and then measure the enhanced rotation-rate transduction sensitivity near the EP. The laser gyroscope is described elsewhere \cite{li2017microresonator} and uses counter-propagating Brillouin lasers in a high-quality-factor ($Q\approx10^8$) silica wedge resonator \cite{lee2012chemically}. As shown in Fig. \ref{Fig1}a pump light at frequencies $\omega_{p_j}$ ($j=1,2$) determined by radio-frequency modulation of a single laser ($\sim$1552.5 nm) is coupled into the resonator from both ends of a fiber taper \cite{cai2000observation,spillane2003ideality}. One of the pump frequencies is Pound-Drever-Hall locked to a resonator mode by feedback control to the laser. The second pump frequency is then varied to affect pump detuning change as described below. The two pump powers are stabilized via power feedback. Brillouin scattering causes a pump photon with frequency $\omega_{p_j}$ to scatter from a co-propagating acoustic phonon with frequency $\Omega_{\textrm{phonon}}$ into a backward-propagating Stokes photon with frequency $\omega_{s_j}$. In the context of a resonator (and as illustrated in Fig. \ref{Fig1}b), the associated phase matching condition requires that the Brillouin shift frequency ($\Omega_{\textrm{phonon}}$) is close in value to a multiple of the resonator free-spectral-range (FSR). This is readily achieved by microfabrication control of resonator diameter and in effect locates a resonator mode (the Stokes mode) within the Brillouin gain spectrum for efficient stimulated Brillouin laser (SBL) action \cite{lee2012chemically,li2012characterization}.

Counter-pumping is performed on the same resonant mode number ($m$) so that laser action on two counter-propagating Stokes waves also occurs on one mode number (set to $m$-6 in this measurement). The device then functions as a ring laser gyroscope in which rotation measurement is possible by comparing (via heterodyne detection) the relative frequencies of the Stokes waves \cite{chow1985ring}. Line narrowing to sub-Hertz values results from laser action within the high-Q system \cite{li2012characterization} thereby boosting the rotational sensitivity. To control the frequency of the Stokes waves for study of EP physics, a small frequency mismatch between the cavity mode and the gain spectral center is used to vary modal dispersion and thereby pull the Stokes lasing modes toward the gain spectral center (Fig. \ref{Fig1}b). As described below this pulling mechanism is arranged to be different for the two Stokes waves and it is controlled by varying the pump frequency detuning $\Delta\omega_p = \omega_{p_2} - \omega_{p_1}$. Finally, because of the taper-induced dissipative coupling (rate $\kappa$) between CW and CCW modes, the EPs emerge for a critical pump detuning $\Delta\omega_p =\Delta\omega_c$. When $\Delta\omega_p$ is close to $\Delta\omega_c$ the gyro sensitivity is boosted due to the steep slope of the square-root response curve (Fig. \ref{Fig1}c).

The above features of the gyro system are now studied quantitatively. For simplicity, the case without rotation is considered first. The equation of motion of the system reads $ id\Psi/dt=H_0\Psi$ where $\Psi=(\alpha_1, \alpha_2)^T$ is the column vector for the two SBL modes. $H_0$ is the non-Hermitian Hamiltonian governing the time evolution:
\begin{equation}
\label{EOM}
H_0
=\begin{pmatrix}
\omega_0+i\left(g_1|A_1|^2-\gamma/2\right) & i\kappa \\
i\kappa & \omega_0+i\left(g_2|A_2|^2-\gamma/2\right)\\
\end{pmatrix}
\end{equation}
and $\alpha_1$ ($A_1$) and $\alpha_2$ ($A_2$) represent the photon-number-normalized amplitudes of the CW and CCW SBL (pump) modes, respectively. $\omega_0$ is the unpumped frequency of the Stokes' cavity mode and $\gamma$ is the cavity damping rate. $g_j={g_0}/({1+{2i\Delta\Omega_j}/{\Gamma}})$ ($j=1,2$) represents the Brillouin gain factor where $g_0$ is the gain coefficient, $\Gamma$ is the gain bandwidth, and $\Delta\Omega_j=\omega_{p_j}-\omega_{s}-\Omega_{\textrm{phonon}}$ is the frequency mismatch with $\omega_s$ the Stokes frequency and $\Omega_{\textrm{phonon}}$ the Brillouin shift \cite{li2012characterization}. The real part of the Brillouin gain factor leads to amplification of the Stokes mode, while the imaginary part is responsible for the mode pulling effect. $\kappa$ is the dissipative coupling rate between two SBL modes, which originates from the fiber taper and imperfections in the resonator (see Method). 

In the absence of backscattering ($\kappa=0$), the CW and CCW SBL processes are independent because the Brillouin gain is intrinsically directional as a result of the phase matching condition (Fig. \ref{Fig1}a). The steady-state lasing condition requires the power loss rate $\gamma$ to be balanced by the Brillouin gain, which leads to the clamping condition of the pump powers $|A_j|^2=\gamma(1+{4\Delta\Omega_j ^2}/{\Gamma^2})/{2g_0}$ \cite{li2012characterization}. 
As shown in the Methods, these conditions remain valid for nonzero dissipative backscattering ($\kappa\not=0$) within the regime where EP-enhanced rotation measurement is performed (the unlocked regime defined below). As a result, Eq. (\ref{EOM}) simplifies above laser threshold to the following form:
\begin{equation}
\label{Hthres}
    H_0=\begin{pmatrix}
    \omega_0+\frac{\gamma}{\Gamma}\Delta\Omega_1 & i\kappa \\
    i\kappa & \omega_0+\frac{\gamma}{\Gamma}\Delta\Omega_2
    \end{pmatrix}
\end{equation}
With the introduction of $\kappa$ the lasing system exhibits a frequency locking-unlocking transition when varying the pump detuning frequency. The locking regime is known in ring laser gyroscopes to create a sensing dead band for rotations \cite{chow1985ring}. In the frequency unlocked regime, the two lasing modes oscillate with distinct angular frequencies $\omega_{s+}$ and $\omega_{s-}$, which are the eigenvalues of the Hamiltonian (Eq. (\ref{Hthres})).
\begin{equation}
\label{nustokes}
\omega_{s\pm}-\omega_r=\frac{\gamma/2\Gamma}{1+\gamma/\Gamma}\left(\Delta\omega_p\pm\sqrt{\Delta\omega_p^2-\Delta \omega_c^2} \ \right)
\end{equation}
where $\omega_r \equiv \omega_0+\gamma(\omega_{p_1}-\Omega_{\textrm{phonon}})/\Gamma$ and $\Delta\omega_c \equiv 2 \Gamma\kappa/\gamma$ is the critical frequency. In deriving this result it is important to note that the Hamiltonian (Eq. (\ref{Hthres})) depends weakly upon its own eigenvalues through the appearance of $\Delta \Omega_1$ and $\Delta \Omega_2$ (see derivation in Methods). The SBL beating frequency is readily extracted by taking the difference of the above eigenfrequencies, $\Delta\omega_s \equiv |\omega_{s+}-\omega_{s-}|$:
\begin{equation}
\label{Eq4}
    \Delta\omega_s =\frac{\gamma/\Gamma}{1+\gamma/\Gamma}\sqrt{\Delta\omega_p^2-\Delta \omega_c^2}
\end{equation} % YH: I have modified the equation according to the feedback.

\noindent This equation is plotted in Fig. \ref{Fig1}c. The dissipative coupling between the clockwise (CW) and counterclockwise (CCW) lasing modes induces second-order EPs at critical pump-detuning frequencies $|\Delta\omega_p|=\Delta\omega_c$ where the eigenfrequencies as well as the eigenmodes coalesce. For pump detuning $|\Delta \omega_p|>\Delta\omega_c$ the eigenfrequencies bifurcate and the eigenmodes are an unbalanced hybridization of CW and CCW modes. For pump detuning $|\Delta \omega_p|<\Delta\omega_c$ the eigenfrequencies (real part of the eigenvalues) are equal, but have different loss rates.

An electrical spectrum analyzer was used to measure the photo-detected dual-SBL beating frequency $\Delta\omega_s / 2 \pi$ (Fig. \ref{Fig2}a) and the SBL-pump beating frequency (Fig. \ref{Fig2}b). Plots of these frequencies versus the pump frequency detuning are given in Fig. \ref{Fig2}c and Fig. \ref{Fig2}d. Comparisons with Eq. (\ref{nustokes}) and Eq. (\ref{Eq4}) are provided and are in good agreement with measurement. Moreover, the ratio of the CCW components in the eigenmodes was measured from the intensity of the CCW-pump beating with the SBL signals (see Method for analysis) and is plotted as the inset of Fig. \ref{Fig2}d. There is a reasonable agreement between the model and measurement. Within the locked regime, only one Stokes mode is lasing so this measurement is no longer possible. Further discussion is provided in the Methods.

\begin{figure}
    \centering
    \includegraphics[width=1\columnwidth]{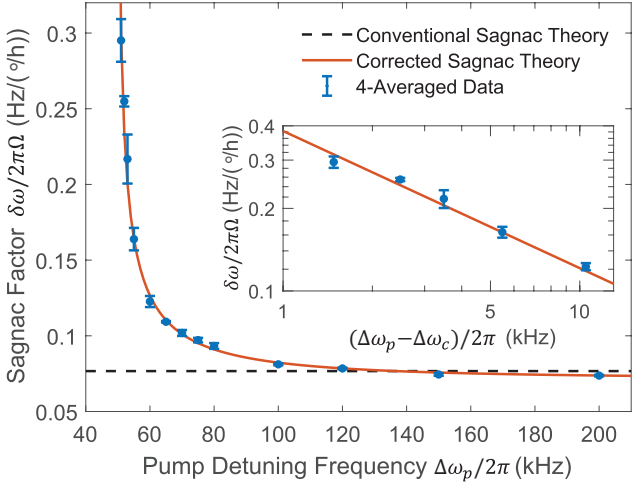}
    \caption{{\bf Measured Sagnac transduction factor $S(\Delta\omega_p)$ compared with model.} The blue dots are data (each point is an average of four measurements) while the red curve is the theoretical prediction using Eq. (\ref{Sagnac}). The mode pulling factor $1/(1+\gamma/\Gamma)$ slightly reduces the Sagnac factor at large pump detuning. The black dashed line gives the conventional (non EP-enhanced) Sagnac factor. The inset shows a log-log plot of 5 data points near the EP with a slope of -1/2, further verifying that the sensitivity enhancement is proportional to $(\Delta\omega_p-\Delta\omega_c)^{-1/2}$.}
    \label{Fig3}
\end{figure}

When the resonator experiences an angular rotation rate $\Omega$ (positive for CW direction), the Sagnac effect further lifts the degeneracy of the CW and CCW modes by shifting the CW and CCW mode frequencies by $\Delta \omega_{\rm Sagnac} = \mp 2 \pi D\Omega/n_g\lambda$ where $D$ is the resonator diameter, $n_g$ is the group index of the passive cavity mode, and $\lambda$ is the laser wavelength \cite{li2017microresonator}. This modifies the SBL beating frequency as follows:
\begin{equation}
\label{Eq5}
    \Delta\omega_s =\frac{\gamma/\Gamma}{1+\gamma/\Gamma}\sqrt{\left(\Delta\omega_p - \Gamma \Delta \omega_{\rm Sagnac}/ \gamma\right)^2-\Delta \omega_c^2}
\end{equation} 
%\begin{equation}
%H_{Sagnac}
%={2\pi}\begin{pmatrix}
%-D\Omega/2n_g\lambda & 0 \\
%0 & D\Omega/2n_g\lambda\\
%\end{pmatrix}
%\end{equation}
%where $\Delta \Omega \omega_{\rm Sagnac} \equiv (2 \pi \Gamma D\Omega) / (\gamma n_g\lambda)$. In this expression, $\Omega$ represents the angular velocity of the rotation, $D$ is the resonator diameter, $n_g$ is the group index of the passive cavity mode, and $\lambda$ is the laser wavelength. 
Accordingly, the counter-pumped Brillouin system can serve as a gyroscope for measuring the rotation signal $\Omega$ by monitoring the dual-SBL beating $\Delta\omega_s$. For comparison with measurements below, the Sagnac transduction factor $S$ is calculated as the derivative of the SBL splitting frequency with respect to the applied rotation rate amplitude $\Omega$:
\begin{equation}
\label{Sagnac}
  {S}=\left.\frac{\partial\Delta\omega_s}{\partial\Omega}\right|_{\Omega=0}=\frac{2 \pi}{1+\gamma/\Gamma}\frac{\Delta\omega_p}{\sqrt{\Delta\omega_p^2-\Delta\omega_c^2}}\frac{D}{n_g\lambda}
\end{equation}
where a linear response requires $\Gamma \Delta \omega_{\rm Sagnac}/ \gamma \ll \Delta\omega_p$. In this equation, the coefficient $1/(1+\gamma/\Gamma)$ is a correction from the mode pulling effect and the factor ${\Delta\omega_p}/{\sqrt{\Delta\omega_p^2-\Delta\omega_c^2}}$ is the transduction enhancement caused by the square-root dependence at the EP. This enhancement originates from the steep slope of the response curve near the EP (Fig. \ref{Fig1}c) so that the Sagnac transduction factor surpasses the conventional value. 

To measure rotations and verify the EP enhancement, the resonator was packaged in a small metal box with one edge hinged and the opposing end attached to a PZT stage in a manner similar to that used in ref. \cite{li2017microresonator}. A sinusoidal oscillation was generated by the PZT to create a sinusoidal rotation of the resonator at a 1 Hz rate having a fixed amplitude (410 deg/h). The resulting time-varying dual-SBL beating frequency was recorded using a frequency counter and the amplitude of the modulated frequency was extracted by applying a fast-Fourier transform to the counter signal. Frequency modulation amplitudes were recorded at a series of pump frequency detunings. The resulting Sagnac transduction factor (i.e., the SBL difference frequency modulation amplitude divided by the applied rotation-rate amplitude) is plotted in Fig. \ref{Fig3}. A boosted transduction factor by up to 4$\times$ compared to the non-EP-enhanced case is observed when operating close to the EP (i.e., near the critical detuning frequency). There is good agreement between Eq. (\ref{Sagnac}) and the measurement as shown in Fig. \ref{Fig3}. As an aside, thermal and pumping power fluctuations were the primary source of noise affecting the measurement. Specifically, as the system operates closer to the EP, these mechanisms exert a greater impact on the measurement leading to relatively larger error bars \cite{EPnoise}.

In summary, exceptional-point-enhancement of a laser gyroscope's sensitivity has been experimentally demonstrated. The gyroscope implementation used a microresonator Brillouin laser and verified a boost of the Sagnac rotation transduction factor up to 4$\times$ near the EP. The microresonator-based counter-pumped Brillouin system makes it possible to engineer the gain/loss rate of CW and CCW modes independently and can facilitate the study of EP physics. This work therefore provides a new platform for studying EPs in a nonlinear optical system while also demonstrating sensitivity improvement of on-chip sensors.

 \renewcommand{\figurename}{\textbf{Extended Data Figure}}

%%%%%%%%%%%%%%%%%%%%%%%%%% Begin Method %%%%%%%%%%%%%%%%%%%%%%%%%%

\begin{comment}
\begin{figure*}
    \centering
    \includegraphics[width=2\columnwidth]{Ext_Fig.eps}
    \caption{
    \textbf{a}, The dual-SBL beating spectrum in the baseband. The high harmonics comes from the Adler solution derived from the coupled-mode theory. \textbf{b}, The pump-SBL beating spectrum in the high frequency ($\approx 10.845$ GHz) centered by the pump1-SBL1 beating peak. The individual pump-SBL beating peaks are identified. The other peaks in the spectrum are either noise from the environment, or high harmonics originate from the nonlinear nature of the Brillouin process.}
    \label{SFig}
\end{figure*}
\end{comment}

\begin{footnotesize}

\section*{Methods}

%\noindent \textbf{Measurement setup} 
%The microresonator ($Q\approx10^8$) had a diameter of $D=36mm$ (FSR $\sim 1.8$GHz) in order to satisfy the Brillouin phase matching condition (see Fig. \ref{Fig1}). A single pump laser with wavelength $\lambda_p=1552.5$nm was split into two arms that were separatedly shifted in frequency by acoustic-optical modulators. The first arm (Pump 1) is Pound-Drever-Hall locked to the cavity mode and the second arm (Pump 2) is shifted to precisely control the pump frequency detuning $\Delta\nu_p=\nu_{p_2}-\nu_{p_1}$. The two pump powers are stabilized via power feedback are set to be balanced in order to remove the Kerr-induced frequency shift (see discussion below). The pump-SBL beating and dual-SBL beating signals are measured by photodetectors that are connected to either an electrical spectrum analyzer or a frequency counter.

\medskip

\noindent \textbf{Origin of the dissipative coupling} 
In a standing-wave mode basis, the optical loss induced by the fiber taper or any other spatially localized absorption or dissipative scattering element will be different for each mode and can be captured by the following contribution to the Hamiltonian:
\begin{equation}
H_{\textrm{taper}}=
   \begin{pmatrix}
    -i\gamma_1 & 0 \\
    0 &-i\gamma_2
    \end{pmatrix} .
\end{equation}
Changing to a traveling wave basis (CW and CCW) by using the relation $\ket{\Phi_\pm}=\left(\ket{\text{CW}}\pm\ket{\text{CCW}}\right / \sqrt{2}$ gives the following Hamiltonian in the new basis,
\begin{equation}
H_{\textrm{taper}}=
   \begin{pmatrix}
    -i\gamma_{\textrm{common}} & 0 \\
    0 &-i\gamma_{\textrm{common}}
    \end{pmatrix} 
    +\begin{pmatrix}
    0 & i\kappa \\
    i\kappa & 0
    \end{pmatrix} 
\end{equation}
where $\gamma_{\textrm{common}}=(\gamma_1+\gamma_2)/2$ and $\kappa=(\gamma_1-\gamma_2)/2$. The first term is the common loss (out-coupling loss of the taper) while the second term is the dissipative backscattering in Eq. (\ref{EOM}).

\medskip

\noindent \textbf{Validity of clamping condition} Note that the Hamiltonian in Eq.\ref{EOM} depends on its eigenvalues $\omega_{s_j}$ through the Brillouin gain factor $g_j=g_0/[1+2i(\omega_{p_j}-\Omega_{\textrm{phonon}}-\omega_{s_j})/\Gamma]$. However, by separating the Brillouin gain factor into real part and imaginary parts as follows:
\begin{eqnarray}
\label{SBSgain1}
\textrm{Re}(g_j)&=&\frac{g_0}{1+4\frac{(\omega_{p_j}-\Omega_{\textrm{phonon}}-\omega_{s_j})^2}{\Gamma^2}}\\
\label{SBSgain2}
\textrm{Im}(g_j)&=&\textrm{Re}(g_j)\left[1-2i\left(\omega_{p_j}-\Omega_{\textrm{phonon}}-\omega_{s_j}\right)/\Gamma\right]
\end{eqnarray}
it can seen that for mode pulling that is small compared to the cavity linewidth (which is the case in this work), $\omega_{s_j}$ can be replaced by $\omega_0$ in the denominators of Eq. (\ref{SBSgain1}) and Eq. (\ref{SBSgain2}) leaving the eigenvalue dependence only in the dispersive term (numerator). Furthermore, by defining normalized quantities:
\begin{widetext}
\begin{eqnarray}
    I_j\equiv \frac{\textrm{Re}(g_j)|A_j|^2}{\gamma/2},\quad k\equiv \frac{\kappa}{\gamma/2},\quad n_{pj}\equiv \frac{\omega_{p_j}-\Omega_\textrm{{phonon}}}{\gamma/2},\quad x\equiv\frac{\omega_s}{\gamma/2},\quad x_0\equiv\frac{\omega_0}{\gamma/2}
    ,\quad r\equiv\frac{\gamma}{\Gamma},
\end{eqnarray}
the Hamiltonian reduces to:
\begin{eqnarray}
\tilde{H_0}\equiv\frac{H_0}{\gamma/2}
={x_0}\mathbb{1}+\begin{pmatrix}
i\left(I_1-1\right)+rI_1(n_{p1}-x) & ik \\
ik & i\left(I_2-1\right)+rI_2(n_{p2}-x)\\
\end{pmatrix}
\end{eqnarray}
The eigenvalues $x_\pm$ can be solved from $\det(\tilde{H_0}-x\mathbb{1})=Ax^2+Bx+C=0$ where
\begin{eqnarray}
A&=&(1 + I_1 r) (1 + I_2 r)\\
B&=&2 i - 2 x_0 -(I_1+ I_2) (i  -i r+ x_0 r) - r(I_1n_{p1}+I_2n_{p2})
 +I_1  I_2[2 i + (n_{p1} + n_{p2})r ]\\
C&=&k^2 + (-i + x_0)^2 + (-i + x_0) [I_1(i + n_{p1} r)+I_2(i + n_{p2} r)] + 
 I_1 I_2(i + n_{p1} r) (i + n_{p2} r)
\end{eqnarray}
Because the two eigenvalues $x_\pm=(-B\pm\sqrt{B^2-4AC})/2$ should both be real (i.e., above laser threshold operation), the following equations can be derived from $\text{Im}(x_\pm)=0$
\begin{eqnarray}
    \text{Im}(B^2-4AC)&=&2 r (r+1) ({I_1}-{I_2}) [{I_1}{I_2} r (n_{p1}-n_{p2})+I_1({np_1}-{x_0})+{I_2} ({x_0}-{n_{p2}})]=0\label{clamp1}\\
  \text{Im}(B)&=& 2r I_1{I_2}+(I_1+{I_2})(1-r)-2=0\label{clamp2}
\end{eqnarray}
\end{widetext}
It can be obtained from Eq. (\ref{clamp1}) that $I_1=I_2$. Inserting this result into Eq. (\ref{clamp2}) gives $I_1=I_2=1$ yielding $|A_j|^2=\gamma(1+{4 \Delta\Omega_j ^2}/{\Gamma^2})/{2g_0}$ where $\Delta\Omega_j = \omega_{p_j}-\Omega_{\textrm{phonon}}-\omega_{s_j}$. These are also the $\kappa = 0 $ gain clamping conditions used to simplify the Hamiltonian to the form given in Eq. (\ref{Hthres}). Numerical solution of the eigenvalue equation confirms that this result holds for the unlocked regime. On the other hand, numerical solution also shows that in the locked regime only one eigenvalue can be real for any combination of pumping powers (i.e., only one mode lases in the locked regime). Moreover, a low and high loss eigenvalue exist so that one mode has a lower threshold pumping power. An equal pump power solution ($I_1=I_2$) is still possible for laser action, but this condition is no longer unique.

\medskip

\newpage

\noindent \textbf{Characterization of eigenmodes} 

%In the absence of rotation, the Hamiltonian (above lasing threshold) in the unlocked regime is obtained by substituting the clamping conditions discussed above into Eq. (\ref{EOM}) yielding:
%\begin{equation}
%    H_0=\begin{pmatrix}
%    \omega_0+\frac{\gamma}{\Gamma}\Delta\Omega_1 & i\kappa \\
%    i\kappa & \omega_0+\frac{\gamma}{\Gamma}\Delta\Omega_2
%    \end{pmatrix}
%\end{equation}
%The eigenvalues are 
%\begin{equation}
%\omega_{s\pm}-\omega_r=\frac{1}{1+\gamma/\Gamma}\left[\frac{\gamma}{2\Gamma}\Delta\omega_p\pm\sqrt{\left(\frac{\gamma}{2\Gamma}\Delta\omega_p\right)^2-\kappa^2}\right],
%\end{equation}
\noindent The eigenmodes of Eq. (\ref{Hthres}) are:
\begin{eqnarray}
    \ket{\Psi_+}&=& \frac{1}{N}\begin{pmatrix}
    -i\\
   \left[{\Delta\omega_p}/{\Delta\omega_c}+\sqrt{\left({\Delta\omega_p}/{\Delta\omega_c}\right)^2-1}\right]
    \end{pmatrix}\\
    \ket{\Psi_-}&=& \frac{1}{N}\begin{pmatrix}
   \left[{\Delta\omega_p}/{\Delta\omega_c}+\sqrt{\left({\Delta\omega_p}/{\Delta\omega_c}\right)^2-1}\right]\\
   i
    \end{pmatrix}
\end{eqnarray}
where $N$ is normalization. These lasing eigenmodes are valid in the uncoupled regime of operation ($\left|\Delta \omega_p\right| > \Delta \omega_c$) and are hybrid modes of the original CW and CCW modes. To make the data plot within the inset of Fig. \ref{Fig2}d the laser output in the CCW direction (combination of two laser Stokes waves) was monitored. This combined CCW field is given by:
\begin{equation}
\label{CCW}
\ket{\text{CCW}}=\frac{1}{N^\prime}\left\{\left[{\Delta\omega_p}/{\Delta\omega_c}+\sqrt{\left({\Delta\omega_p}/{\Delta\omega_c}\right)^2-1}\right]\ket{\Psi_-}-i\ket{\Psi_+}\right\}
\end{equation}
Where $N^\prime$ is another normalization. The ratio of powers of the components was determined by heterodyning this field with a CCW pump field and then measuring the respective Pump-SBL$_{1,2}$ beat components on an electrical spectrum analyzer. The ratio of the powers in these beat frequency components is the ratio of the powers in the CCW Stokes' waves components:
\begin{equation}
    \frac{I_{s2}}{I_{s1}}=\left|\left|{\Delta\omega_p}/{\Delta\omega_c}\right|+\sqrt{\left({\Delta\omega_p}/{\Delta\omega_c}\right)^2-1}\right|^2
\end{equation}
which directly follows from Eq. (\ref{CCW}).

%It is also possible to express the eigenvectors by defining a parameter $r$ where $\Delta\omega_p/\Delta\omega_c=\cosh(2r)$. 
%\begin{equation}
%\label{Eigenvec}
%    \ket{\Psi_\pm}=\frac{1}{\sqrt{2\sinh(2r)}}\begin{pmatrix}
%    e^{\mp r}\\
%   ie^{\pm r}
%    \end{pmatrix}
%\end{equation}
%In the unlocked regime ($\Delta\omega_p/\Delta\omega_c > 1$)  $r>0$. 

It is also interesting to note that in the locked regime ($\left|\Delta\omega_p\right|/\Delta\omega_c < 1$) numerical solution shows that eigenvector solutions having equal admixture of CW and CCW waves occur when $I_1=I_2$, but at distinctly different threshold power levels (i.e., the two states have different loss rates). Moreover, this pumping combination is not unique so lasing solutions featuring an unbalanced admixture of CW and CCW states are also possible. The (locked regime) equatorial trajectories shown in Fig. \ref{Fig1} represent the low and high loss $I_1=I_2$ trajectories (i.e., equal CW and CCW admixture).

As an aside, the measurement in Fig. \ref{Fig2}c and Fig. \ref{Fig2}d use the beat note spectra in Fig. \ref{Fig2}a and Fig. \ref{Fig2}b. There are additional lines in these spectra that are believed to originate from nonlinear mixing in the Brillouin interaction (a third order nonlinear interaction). This four-wave-mixing process becomes more significant near the EP where the CW and CCW modes strongly interact with each other. It impacts the intensity of the beating lines but leaves their frequencies intact. As a result, data for the eigenmode components slightly fluctuate around the theoretical value while the data of the pump-SBL and dual-SBL frequencies fit well with the theory (see Fig. \ref{Fig2}c and d).

\medskip

\noindent \textbf{Kerr-induced shift}
The Kerr effect shifts the resonance frequency by adding the following term into the Hamiltonian:
\begin{equation}
H_\text{Kerr}=
   \begin{pmatrix}
    \eta\left(\left|\alpha_1\right|^2+2\left|\alpha_2\right|^2\right) & 0 \\
    0 &\eta\left(2\left|\alpha_1\right|^2+\left|\alpha_2\right|^2\right)
    \end{pmatrix} 
\end{equation} 
where $\eta=n_2\hbar\omega^2c/Vn_0^2$ is the single photon induced nonlinear angular frequency shift. The corrected beating frequency (without rotation) reads: 
\begin{equation}
    \Delta\omega_s=\frac{1}{1+\gamma/\Gamma}
    \sqrt{\left[\frac{\gamma}{\Gamma}\Delta\omega_p+{\eta}\left(\left|\alpha_1\right|^2-\left|\alpha_2\right|^2\right)\right]^2-4\kappa^2}
\end{equation}
The correction from the Kerr effect is therefore equivalent to shifting $\Delta\omega_p$ by angular frequency $\eta\Gamma(|\alpha_1|^2-|\alpha_2|^2)/\gamma$. In the experiment, this Kerr shift was minimized by centering the locking zone at zero pump detuning by adjusting the two pump powers. After that, the pump powers were locked so that the two SBL powers are balanced. The subsequent pump detuning changes required to make the measurement affected the SBL power, but only negligibly. Specifically, the Kerr shift is around 10s of Hz after a pump detuning change by 200kHz. This is negligible in comparison to the Stokes frequency separation changes measured in Fig. \ref{Fig2}c and Fig. \ref{Fig2}d. Moreover, the dithering measurement in Fig. \ref{Fig3} was insensitive to these constant Kerr-induced shifts since it measured the amplitude of a sinusoidal rotation.

\end{footnotesize}

%%%%%%%%%%%%%%%%%%%%%%%%%% End Method %%%%%%%%%%%%%%%%%%%%%%%%%%

\bibliography{Bib}

\vbox{}
\noindent \textbf{Acknowledgments} We thank Mercedeh Khajavikhan, Demetrios Christodoulides, Or Peleg, and Barry Loevsky for the helpful discussions in preparing this manuscript. We also thank Boqiang Shen, Chengying Bao, and Qifan Yang for technical support. This project was supported by the Defense Advanced Research Projects Agency (DARPA) through SPAWAR (N66001-16-1-4046) and the Kavli Nanoscience Institute. 

\vbox{}
\noindent \textbf{Author Contributions} Y.-K. L., Y.-H. L., M.-G. S. and K. V. conceived the exceptional point enhancement in the offset-counter-pumped stimulated Brillouin laser gyroscope. Y.-K. L., Y.-H. L. and K.V. constructed the theoretical model. M.-G. S. fabricated the ultra-high-Q silica microresonator and help Y.-H. L. with the packaging. Y.-H. L. and Y.-K. L. performed the experiment. All authors analyzed the data and wrote the manuscript.

\end{document}